\documentclass[%
 reprint,
 amsmath,amssymb,
 aps,
pra,
]{revtex4-2}

\usepackage{graphicx}
\usepackage{dcolumn}
\usepackage{tikz}
\usetikzlibrary{calc}
\usepackage{bm}

\newcommand{\dbar}{\text{d}\hspace*{-0.15em}\bar{}\hspace*{0.1em}}

\usepackage{physics}

\usepackage{hyperref}
\hypersetup{
    colorlinks = true,
    citecolor = blue,
    urlcolor = blue
}

\begin{document}

\title{Trade-off between coherence and heat in a non-Markovian dephasing dynamics}

\author{Marino P. Lenzarini}
\email{marinopl@usp.br}

\author{Diogo O. Soares-Pinto}%
 \email{dosp@ifsc.usp.br}
\affiliation{%
Instituto de Física de São Carlos, Universidade de São Paulo, CP 369, 13560-970, São Carlos, SP, Brasil
}%


\begin{abstract}
How quantum coherence influences thermodynamic behavior remains an open question in quantum thermodynamics. Here we investigate this relation within the pure dephasing framework, where a central qubit interacts with a finite Ising-like spin environment. Although the system's internal energy remains constant, the interaction induces decoherence and gives rise to nontrivial thermodynamic features. Within the two-point measurement approach, we show that the heat dissipated into the environment matches the coherent energy contribution appearing in a reformulated first law of quantum thermodynamics. Numerical calculations reveal oscillatory coherence dynamics, with revivals associated with information backflow and non-Markovian effects, as quantified by the Breuer–Laine–Piilo measure. We find that heat and coherence exhibit intertwined temporal behavior, with enhanced heat dissipation during coherence decay and reduced heat during revivals. These results suggest a connection between coherence dynamics and thermodynamic quantities in finite, closed composite systems undergoing pure dephasing.

\end{abstract}

\maketitle


\section{\label{sec1}Introduction}

Understanding the thermodynamic cost of quantum coherence has become an important question at the interface between quantum information theory and quantum thermodynamics~\cite{campbell2025roadmap, goold2016role, hammam2022exploiting, su2018heat, francica2019role, lostaglio2015description, korzekwa2016extraction}. In particular, when open quantum systems are considered, the unavoidable interaction between the system of interest and its environment leads to decoherence, which may accompany thermalization depending on the coupling strength and spectral properties of the environment. In many settings, equilibration proceeds through energy exchange with the environment, promoting the emergence of classical behavior~\cite{schlosshauer2005decoherence, schlosshauer2007decoherence, zurek1991decoherence, rivas2012open, breuer2002theory, joos1985emergence, chen2019quantifying}.

Among the theoretical frameworks employed to investigate decoherence, pure dephasing dynamics provides a simple yet paradigmatic setting for exploring how quantum coherence and thermodynamic quantities are related. In this regime, the energy of the open quantum system is conserved and thermalization is inhibited, while coherence decays solely due to system–environment correlations. This framework has served as a benchmark for understanding the time-dependent entanglement between a qubit and its environment, providing a suitable setting for the application of spin-echo protocols~\cite{roszak2015qubit, roszak2015characterization}. It has also been employed to analyze the backflow of quantum information in different scenarios, including a qubit coupled to an Ising spin bath~\cite{krovi2007nonmarkovian}, bosonic environments~\cite{xiong2020boson}, and spin-ensemble models~\cite{dubertrand2018analytical}, offering a controlled platform to investigate non-Markovian features.

At first sight, since the internal energy of the open system remains constant, pure dephasing is often regarded as energetically trivial. However, it has been shown that even though the system energy remains conserved, nontrivial heat dissipation can emerge as a redistribution of energy involving the interaction term, while the total energy of the composite system remains conserved~\cite{popovic2023thermodynamics}. This observation opens the possibility of investigating how quantum coherence is intrinsically connected to thermodynamic quantities in such scenarios.

Recent developments in quantum thermodynamics have further clarified that coherence plays a central role in the variation of the internal energy of a quantum system. In particular, Bernardo~\cite{bernardo2020unraveling} proposed an extension of the first law, formulated within the two-point measurement (TPM) scheme, in which an additional term — referred to as coherent energy — accounts explicitly for a genuinely quantum contribution without classical analog, capturing the energetic role of quantum coherence.

In this work, we show analytically that, although the internal energy of the open system remains constant and the total energy is conserved, the mean heat dissipated into the environment — defined within the TPM framework — coincides exactly with the coherent energy contribution appearing in Bernardo’s reformulated first law. This finding motivates us to investigate numerically the thermodynamic aspects of pure dephasing dynamics in finite environments, providing evidence of an intertwined dynamical behavior between dissipated heat and coherence loss and recovery.

To illustrate this connection in a concrete setting, we consider a central qubit coupled simultaneously to all spins in a finite Ising-like spin environment arranged in a ring configuration. The finiteness of the environment induces coherence revivals and information backflow, allowing us to connect dissipated heat, coherence dynamics, and non-Markovian behavior in a single setting. Numerical calculations reveal that maxima of the mean heat correspond to minima of the $l_1$-norm of coherence of the qubit, while coherence revivals coincide with reductions in heat, reflecting the reversible exchange of quantum information between system and environment.

The paper is organized as follows. In Sec.~\ref{sec2} we present the general framework of pure dephasing dynamics and analyze how decoherence emerges. In Sec.~\ref{sec3} we examine its thermodynamic structure and establish the equivalence between mean heat and coherent energy based on Bernardo's proposal. In Sec.~\ref{sec4} we introduce the physical model and investigate coherence, non-Markovianity, and heat dynamics numerically. Finally, Sec.~\ref{sec5} summarizes our conclusions and outlines directions for future research.

\section{\label{sec2} Pure Dephasing}

Consider an open quantum system, $S$, interacting with a thermal environment, $E$, at temperature $T$. The composite system evolves under the total Hamiltonian
\begin{equation}
    \hat{H} = \hat{H}_S + \hat{H}_E + \hat{H}_I,
\end{equation}
where $\hat{H}_I$ is the interaction term. Pure dephasing dynamics emerges when the system Hamiltonian is a constant of motion during the evolution~\cite{roszak2015qubit, popovic2023thermodynamics}. Therefore, $\hat{H}_S$ must commute with the interaction Hamiltonian, leading to the pure dephasing condition
\begin{equation}
    \label{pure dephasing condition}
    [\hat{H}_S, \hat{H}_I] = 0.
\end{equation}

The most general form of $\hat{H}_I$ that satisfies this condition is
\begin{equation}
    \hat{H}_I = \sum_k \ketbra{k} \otimes g_k \hat{V}_k,
\end{equation}
where $\{\ket{k}\}$ are the eigenstates of the system Hamiltonian constructed via its spectral decomposition as
\begin{equation}
    \label{system spectral decomposition}
    \hat{H}_S = \sum_k \epsilon_k \ketbra{k},
\end{equation}
with associated eigenvalues $\{\epsilon_k\}$, $\hat{V}_k$ are arbitrary operators acting on the environment, and $g_k$ are coupling constants that characterize the interaction strength between the environment and each eigenstate of $S$.

The initial composite system is taken to be uncorrelated prior to the interaction, $\hat{\rho}(0) = \hat{\rho}_S(0) \otimes \hat{\rho}_E(0)$, where $\hat{\rho}_E(0) = e^{-\beta\hat{H}_E}/Z_E$ is the thermal Gibbs state of the environment, with $\beta = T^{-1}$ denoting the inverse temperature (we use units where $k_B = \hbar = 1$), and $Z_E = \tr_E[e^{-\beta \hat{H}_E}]$ is the partition function.

We assume that the thermal environment is finite and the joint system evolves under unitary dynamics; therefore, the joint state interacts over a period of time $t$ and is given by $\hat{\rho}(t) = \hat{\mathcal{U}}(t)\,\hat{\rho}(0)\,\hat{\mathcal{U}}^{\dagger}(t)$, where the unitary operator is
\begin{equation}
    \hat{\mathcal{U}}(t) = \sum_k \ketbra{k} \otimes \hat{\omega}_k(t),
\end{equation}
with $\hat{\omega}_k(t) = \exp(-i\epsilon_k t)\,\exp(-i\hat{H}_k t)$, in which the first term describes the free phase evolution generated by the system Hamiltonian, whereas the second corresponds to the environment dynamics conditioned on the system state $\ket{k}$, with an effective Hamiltonian given by
\begin{equation}
    \hat{H}_k = \hat{H}_E + g_k\hat{V}_k.
\end{equation}

Since we are interested in the open system behavior, we describe its dynamics through the reduced density operator as $\hat{\rho}_S(t) = \tr_E[\hat{\rho}(t)]$. Thus, we project onto the eigenstates of $\hat{H}_S$ yielding the system's matrix elements
\begin{equation}
    \label{system's elements matrix}
    \rho_S^{k,j}(t) = \rho_S^{k,j}(0)\, \tr_E[\hat{\omega}_k(t)\,\hat{\rho}_E(0)\, \hat{\omega}^\dagger_j(t)].
\end{equation}
As a characteristic feature of pure dephasing, the initial populations of $S$ remain unchanged, and the factor $\tr_E[\hat{\omega}_k(t)\,\hat{\rho}_E(0)\, \hat{\omega}^\dagger_j(t)]$ encodes all information about the environment's influence on the system~\cite{popovic2023thermodynamics}. This motivates the definition of the coherence function
\begin{equation}
    \label{environment coherence}
    \Gamma_{k,j}(t) = \tr_E[\hat{\omega}_k(t)\,\hat{\rho}_E(0)\, \hat{\omega}^\dagger_j(t)].
\end{equation}

One may observe that $\Gamma_{k,j}(t)$ remains constant and equal to unity for $k=j$, as can be verified using the cyclic property of the trace; consequently, the diagonal elements of Eq.~\eqref{system's elements matrix} remain constant during the interaction. Therefore, it is straightforward to apply the $l_1$-norm of coherence as a resource measure~\cite{baumgratz2013quantifying},
\begin{equation}
    \label{l1 norm}
    C_{l_1}(\hat{\rho}_S(t)) = \sum_{k\neq j} |\hat{\rho}_S^{k,j}(0)|\, |\Gamma_{k,j}(t)|,
\end{equation}
which shows that the decoherence process depends explicitly on the magnitude of $\Gamma_{k,j}(t)$ and occurs whenever $\hat{H}_k \neq \hat{H}_j$, indicating that the environment dynamics becomes conditioned on the system state and thus carries information about it.


\section{\label{sec3}Thermodynamic Aspects}
\subsection{Nontrivial Heat Dissipation}

Having developed a comprehensive understanding of the features characterizing pure dephasing and the condition for decoherence, we now turn to its thermodynamic aspects~\cite{lostaglio2015description, francica2019role, korzekwa2016extraction}. To begin, we recall that the internal energy of a quantum system $\hat{\rho}$, governed by the time-dependent Hamiltonian $\hat{H}(t)$, can be evaluated as the expectation value of its Hamiltonian at a given time $t$ as
\begin{equation}
    U(t) = \tr[\hat{H}(t)\,\hat{\rho}(t)].
\end{equation}
Therefore, the infinitesimal variation is
\begin{equation}
    \label{first law classical interpretation}
    \dd U(t) = \tr[\dd\hat{H}(t)\, \hat{\rho}(t)] + \tr[\hat{H}(t)\, \dd\hat{\rho}(t)].
\end{equation}
Following the classical interpretation, the first term is associated with work, understood as the change in internal energy induced by controllable variations in the system's parameters, incorporated as the infinitesimal changes into the Hamiltonian. The second term corresponds to changes in the occupation probabilities encoded in the instantaneous density operator $\hat{\rho}(t)$, which is associated with heat~\cite{filippetto2025convite}.

Within the TPM formulation~\cite{talkner2007fluctuation, talkner2020colloquium}, the change in the internal energy of a composite system, formed by the open quantum system, $S$, and the environment, $E$, is determined from the difference between the outcomes of projective energy measurements performed at the initial, $t=0$, and the final, $t=t_f$, times of the interaction protocol, yielding $\Delta U_{SE} = \langle \hat{H}\rangle_{t_f} - \langle \hat{H}\rangle_0$, where $\langle \bullet \rangle_t = \tr[\bullet\hat{\rho}(t)]$ denotes the expectation value at time $t$.

Since the global dynamics is unitary, the total energy is conserved, implying $\Delta U_{SE} = 0$. Furthermore, the pure dephasing condition in Eq.~\eqref{pure dephasing condition} ensures that the system Hamiltonian is conserved during the evolution, leading to $\Delta U_S = \langle \hat{H}_S\rangle_{t_f} - \langle \hat{H}_S\rangle_0 = 0$. 

Despite this conservation of energy, nontrivial thermodynamic contributions arise. Expanding the expectation value of the total Hamiltonian yields the energy balance
\begin{equation}
    \label{energetic balance}
    -[\langle \hat{H}_I\rangle_{t_f} - \langle \hat{H}_I\rangle_0] = \langle \hat{H}_E\rangle_{t_f} - \langle \hat{H}_E\rangle_0.
\end{equation}
The left-hand side corresponds to the mean work $\langle W \rangle$ required to switch on and off the interaction between $S$ and $E$, while the right-hand side represents the variation of the environment's internal energy, which is identified as the mean heat~\cite{esposito2010entropy}
\begin{equation}
    \label{mean heat definition}
    \langle Q \rangle = \langle \hat{H}_E\rangle_{t_f} - \langle \hat{H}_E\rangle_0,
\end{equation}
leading to the first law of quantum thermodynamics for pure dephasing as $\langle Q \rangle = \langle W \rangle$~\cite{popovic2023thermodynamics}. Importantly, for work and heat to be simultaneously measurable within the TPM framework, the interaction energy must be negligible at the initial and final times of the protocol~\cite{talkner2016fluctuation}.

To confirm that the heat dissipation is nontrivial, i.e., that $\langle Q \rangle$ is indeed nonzero, we construct the reduced state of the environment from the joint state at time $t$ in the system eigenbasis, given by
\begin{equation}
    \hat{\rho}_{SE}(t) = \sum_{k,j} \rho_{S}^{k,j}(0) \ket{k}\bra{j} \otimes \hat{\omega}_k(t)\,\hat{\rho}_E(0)\, \hat{\omega}^\dagger_j(t),
\end{equation}
where $\rho_S^{k,j}(0) = \mel{k}{\hat{\rho}_S(0)}{j}$. By tracing out the system degrees of freedom, we obtain the reduced state of the environment,
\begin{equation}
    \label{environment's reduced state}
    \hat{\rho}_E(t) = \sum_k \rho_S^{k,k}(0)\, \hat{\omega}_k(t)\,\hat{\rho}_E(0)\, \hat{\omega}^\dagger_k(t).
\end{equation}
Substituting into the mean heat definition in Eq.~\eqref{mean heat definition}, we obtain
\begin{align}
    \label{explicit mean heat}
    \langle Q \rangle  =  \sum_k \rho_S^{k,k}(0) \, \tr_E[\hat{H}_E\,\hat{\omega}_k(t)\,\hat{\rho}_E(0)\, \hat{\omega}^\dagger_k(t)] \nonumber\\
    - \tr_E[\hat{H}_E\,\hat{\rho}_E(0)].
\end{align}

Importantly, Eq.~\eqref{explicit mean heat} shows that heat dissipation arises whenever the operators $\hat{\omega}_k(t)$ fail to commute with $\hat{H}_E$; equivalently, in terms of the effective interaction Hamiltonian, a sufficient condition for nonzero dissipation is~\cite{popovic2023thermodynamics}
\begin{equation}
    \label{heat dissipation condition}
    [\hat{H}_E, \hat{H}_k] \neq 0.
\end{equation}
In addition, one may observe that only the diagonal elements of $S$ contribute in Eq.~\eqref{explicit mean heat}, reflecting the fact that energy is conserved in the system while the environment's energy is not conserved in general, as a feature of pure dephasing.


\subsection{Heat as a Signature of Decoherence}

As an alternative way to understand how nontrivial heat is dissipated in this regime, we recall a reformulation of the first law of quantum thermodynamics proposed by Bernardo in~\cite{bernardo2020unraveling}, given by
\begin{equation}
    \dd U = \dbar W + \dbar \mathcal{Q} + \dbar \mathcal{C},
\end{equation}
where the total infinitesimal change in the energy of a quantum system can be decomposed into three distinct contributions: two associated with classical-like thermodynamic processes, such as work, $\dbar W$, and heat, $\dbar \mathcal{Q}$, and a purely quantum-mechanical contribution due to coherence dynamics, $\dbar \mathcal{C}$, referred to as coherent energy.

\subsubsection{Work}

To begin, we focus on the environment's point of view, which is considered the working substance, whose time-independent Hamiltonian is given by
\begin{equation}
    \hat{H}_E = \sum_n \mathcal{E}_n \ketbra{r_n}.
\end{equation}
Then, we evaluate the infinitesimal change in its internal energy as
\begin{equation}
    \label{classical work decomposition}
    \dd U_E = \sum_n[\mathcal{E}_n \, \dd P_n + P_n\,\dd\mathcal{E}_n],
\end{equation}
where $\mathcal{E}_n = \mel{r_n}{\hat{H}_E}{r_n}$ is the energy measured when the environment is in the state $\ket{r_n}$ and $P_n(t) = \mel{r_n}{\hat{\rho}_E(t)}{r_n}$, with $\hat{\rho}_E(t)$ given by Eq.~\eqref{environment's reduced state}, is the corresponding occupation probability. From the classical interpretation (see Eq.~\eqref{first law classical interpretation}), we identify work as
\begin{equation}
    \dbar W = \sum_n P_n\,\dd\mathcal{E}_n,
\end{equation}
associated with changes in the environment's energy levels, and heat as the term associated with changes in the occupation probabilities,
\begin{equation}
    \dbar Q = \sum_n \mathcal{E}_n\,\dd P_n(t).
\end{equation}
In our scenario, $\hat{H}_E$ is time independent, so $\mathcal{E}_n$ does not change in time. Hence, no work arises, in the classical sense, $\dbar W = 0$, and we find that the heat equals the change in the environment's internal energy, $\dd U_E = \dbar Q$, in agreement with Eq.~\eqref{mean heat definition}.

\subsubsection{Heat}

Now, from another perspective, we use the eigenstate basis of the environment's reduced state after the interaction to evaluate the change in the environment's internal energy as
\begin{equation}
    \dd U_E = \sum_{k,m}\left[\dd (\rho_S^{k,k}\rho_E^{m})\,\varepsilon_{m}^{k}(t) 
    + (\rho_S^{k,k}\rho_E^{m})\,\dd \varepsilon_{m}^{k}(t)\right],
\end{equation}
where $\rho_S^{k,k} = \mel{k}{\hat{\rho}_S(0)}{k}$, $\rho_E^m = \mel{r_m}{\hat{\rho}_E(0)}{r_m}$, and $\varepsilon_m^k(t) = \mel{r_m^k(t)}{\hat{H}_E}{r_m^k(t)}$, with $\ket{r_m^k(t)} = \hat{\omega}_k(t)\ket{r_m}$ denoting the rotated environment eigenstate after the interaction protocol. It is straightforward to identify the heat contribution, based on classical concepts, as
\begin{equation}
    \dbar \mathcal{Q} = \sum_{k,m} \dd(\rho_S^{k,k}\rho_E^{m})\,\varepsilon_{m}^{k}(t),
\end{equation}
and the complementary contribution stemming from the interaction between the system and environment as
\begin{equation}
    \dbar \mathcal{W} = \sum_{k,m} (\rho_S^{k,k}\rho_E^{m})\,\dd \varepsilon_{m}^{k}(t).
\end{equation} 
In pure dephasing, the system populations $\rho_S^{k,k}$ are constant and, by construction, the initial environment populations $\rho_E^m$ are time independent; hence, $\dd(\rho_S^{k,k}\rho_E^m)=0$. Therefore, no heat, in the classical sense, arises. Consequently, the change in the environment's internal energy is entirely provided by the interaction term, $\dd U_E = \dbar \mathcal{W}$.

Although this result may seem to contradict the fact that the change in the internal energy equals the mean heat (see Eq.~\eqref{mean heat definition}), in pure dephasing, the work required to couple the system to the environment is entirely converted into heat in the environment upon decoupling at the end of the protocol, within the TPM framework, as can be seen from the energy balance relation in Eq.~\eqref{energetic balance}~\cite{popovic2023thermodynamics}.

\subsubsection{Coherent Energy}

Finally, to elucidate the role of coherence in this reformulation of the first law of quantum thermodynamics, we note that no contributions based on classical interpretations survive in pure dephasing, since $\dbar W = \dbar \mathcal{Q} = 0$. Therefore, focusing on the decomposition in Eq.~\eqref{classical work decomposition}, one finds that
\begin{equation}
    \dd U_E = \dbar Q = \sum_n \mathcal{E}_n \dd \left(\mel{r_n}{\hat{\rho}_E(t)}{r_n}\right).
\end{equation}
Using the environment's reduced state in Eq.~\eqref{environment's reduced state} and the thermal Gibbs state definition, one obtains
\begin{align}
    \dbar Q = \sum_k \sum_{m,n} \rho_S^{k,k}\, \rho_E^m\, \mathcal{E}_n\, \dd\left[\big|c_{m,n}^{k}(t)\big|^{2}\right],
\end{align}
where $c_{m,n}^k(t) = \mel{r_n}{\hat{\omega}_k(t)}{r_m}$ are the transition amplitudes between environment eigenstates induced by the eigenstate $\ket{k}$ of the system. The variation of $\big|c_{m,n}^{k}(t)\big|^{2}$ is nonzero only if $\hat{\omega}_k(t)$ rotates the environment eigenstates, and thus it quantifies a purely quantum contribution associated with coherence dynamics, which is referred to as coherent energy~\cite{bernardo2020unraveling}
\begin{equation}
    \dbar \mathcal{C} = \sum_k \sum_{m,n} \rho_S^{k,k}\, \rho_E^m\, \mathcal{E}_n\, \dd\left[\big|c_{m,n}^{k}(t)\big|^{2}\right].
\end{equation}

Finally, to connect the nontrivial heat dissipation in pure dephasing with the coherent energy, we compute the finite-time coherent energy as
\begin{align}
    \mathcal{C}(t) &= \sum_k \sum_{m,n} \rho_S^{k,k}\, \rho_E^m\, \mathcal{E}_n\,\int_0^t \dv{}{t^\prime}\, \big|c_{m,n}^{k}(t^\prime)\big|^{2}\, \dd t^\prime \nonumber \\ 
    &= \sum_k \sum_{m,n} \rho_S^{k,k}\, \rho_E^m\, \mathcal{E}_n\, \left(\big|c_{m,n}^{k}(t)\big|^{2} - \big|c_{m,n}^{k}(0)\big|^{2}\right),
\end{align}
where, by definition, $\big|c_{m,n}^{k}(0)\big|^{2} = \delta_{m,n}$. Then,
\begin{align}
    \mathcal{C}(t) &= \sum_k \sum_{m,n} \rho_S^{k,k}\, \rho_E^m\, \mathcal{E}_n\, \left(\big|c_{m,n}^{k}(t)\big|^{2} - \delta_{m,n}\right) \nonumber \\
    &= \sum_k \sum_{m} \rho_S^{k.k}\, \rho_E^m\, \left[\sum_n \mathcal{E}_n\big|c_{m,n}^{k}(t)\big|^{2} - \mathcal{E}_m\right].
\end{align}
Rearranging the first term inside the brackets, we obtain
\begin{equation}
    \sum_n \mathcal{E}_n\big|c_{m,n}^{k}(t)\big|^{2} = \mel{r_m}{\hat{\omega}_k(t)\hat{H}_E\,\hat{\omega}_k^\dagger(t)}{r_m},
\end{equation}
and we arrive at the coherent energy given by
\begin{align}
    \label{coherent energy}
    \mathcal{C}(t) = \sum_k \rho_S^{k,k}\, \tr_E\!\left[\hat{H}_E\, \hat{\omega}_k(t)\, \hat{\rho}_E(0)\, \hat{\omega}_k^\dagger(t)\right] \nonumber\\ -\tr\!\left[\hat{H}_E \hat{\rho}_E(0)\right],
\end{align}
which coincides exactly with the mean heat dissipation given by Eq.~\eqref{explicit mean heat}. This result highlights that, in pure dephasing dynamics, the dissipated heat is entirely of quantum origin and serves as a thermodynamic signature of decoherence.

\section{\label{sec4} Central Qubit Decoherence in an Ising-Like Environment}
\subsection{Description of the Model}

\begin{figure}[t]
\centering
\resizebox{\columnwidth}{!}{%
\begin{tikzpicture}[line cap=round, line join=round]
\usetikzlibrary{calc,arrows.meta}

\def\a{4.2}
\def\b{2.0}
\def\ha{0.8}
\def\hb{0.6}
\def\Rspin{0.20}
\pgfmathsetmacro{\Dspin}{2*\Rspin}

\def\angA{275}
\def\angB{325}
\pgfmathsetmacro{\angAplus}{\angA+360}
\pgfmathsetmacro{\angM}{(\angA+\angB)/2}

\tikzset{
  ring/.style={draw=black, dashed, line width=0.9},
  haloOut/.style={fill=red!30, opacity=0.35},
  haloIn/.style={fill=white},
  spinball/.style={
    circle,
    draw=black!65,
    line width=0.35,
    shading=radial,
    inner color=black!10,
    outer color=black!80
  },
  qubit/.style={
    circle,
    draw=blue!55!black,
    line width=0.5,
    shading=radial,
    inner color=blue!10,
    outer color=blue!70
  },
  gline/.style={draw=black!55, dashed, line width=0.9},
  JbondBoth/.style={draw=red!80, line width=2.1,
    {Latex[length=2.4mm]}-{Latex[length=2.4mm]}},
  Jtext/.style={red!80, font=\Large},
  betatext/.style={red!70!black, font=\Large},
  gtext/.style={black!70, font=\Large},
  hzarrow/.style={draw=black!80, line width=1.0, -{Latex[length=2.4mm]}},
hztext/.style={black!80, font=\Large}
}

\path[haloOut] (0,0) ellipse ({\a+\ha} and {\b+\hb});
\path[haloIn]  (0,0) ellipse ({\a-\ha} and {\b-\hb});

\node[betatext]
  at ({(\a+\ha+0.3)*cos(330)},{(\b+\hb-1.9)*sin(330)}) {$\beta$};

\draw[ring]
  ({\a*cos(\angB)},{\b*sin(\angB)})
  arc[start angle=\angB, end angle=\angAplus, x radius=\a, y radius=\b];

\node[qubit, minimum size=1.25cm] (Q) at (0,0) {}; 

\foreach \ang [count=\k] in {25, 75, 125, 175, 225, 275, 325} {
  \path ({\a*cos(\ang)},{\b*sin(\ang)}) coordinate (S\k);
}

\foreach \k in {1,...,7} {
  \ifnum\k=4
    \draw[gline] (Q) -- (S\k) node[midway, below, gtext] {$g$};
    \draw[hzarrow] (S\k) -- ++(0,2.5)
    node[hztext, right] {$h_z$};
    \draw[black!80, line width=1.0] (S\k) -- ++(0,-1.5);
  \else
    \draw[gline] (Q) -- (S\k);
  \fi
}

\foreach \k in {1,...,7} {
  \node[spinball, minimum size=\Dspin cm] at (S\k) {};
}

\draw[JbondBoth]
  ({\a*cos(\angA)},{\b*sin(\angA)})
  arc[start angle=\angA, end angle=\angB, x radius=\a, y radius=\b];

\node[Jtext]
  at ({(\a+0.15)*cos(\angM)},{(\b+0.15)*sin(\angM)-0.20}) {$J_z$};

\draw[black!90, line width=0.75] (-0.20, 0.13) -- (0.20, 0.13); \draw[black!90, line width=0.75] (-0.20,-0.35) -- (0.20,-0.35); \node[font=\scriptsize, black!90] at (0, 0.28) {$|1\rangle$}; \node[font=\scriptsize, black!90] at (0,-0.17) {$|0\rangle$};

\end{tikzpicture}%
}
\caption{
(Color online) Schematic representation of a central qubit (blue sphere) interacting with an environment composed of $N$ spin-$1/2$ particles (black spheres) arranged in a ring geometry. The red halo indicates that the environment is initially prepared in a thermal equilibrium state at inverse temperature $\beta = T^{-1}$. The bath spins interact via a nearest-neighbor Ising coupling of strength $J_z$, and each spin is subject to a local magnetic field $h_z$. The qubit–environment interaction is characterized by the coupling constant $g$.
}
\label{fig:ising_model}
\end{figure}
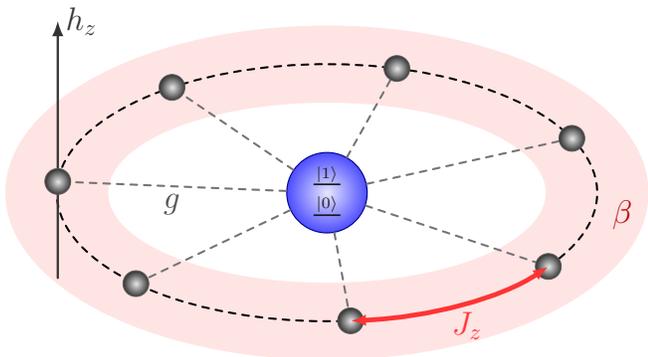

To exemplify how the heat dissipated in a pure dephasing scenario arises as a signature of decoherence, we now consider a physical model consisting of a central qubit coupled to a thermal environment modeled as a finite Ising-like spin chain arranged in a ring configuration, as illustrated schematically in Fig.~\ref{fig:ising_model}. 

Consider the qubit initially prepared in the pure state $\hat{\rho}_S(0) = \ketbra{\psi}$, where $\ket{\psi} = (\ket{0} + \ket{1})/\sqrt{2}$, with its Hamiltonian as defined in Eq.~\eqref{system spectral decomposition}. Now, focusing on the spin chain, the environment Hamiltonian for $N$ spin-$1/2$ particles is given by
\begin{equation}\label{environment's hamiltonian}
    \hat{H}_E = -J_z \sum_{j=1}^N \sigma_j^z \sigma_{j+1}^z - h_z \sum_{j=1}^N \sigma_j^z,
\end{equation}
where $\sigma_j^z$ denotes the Pauli operator acting on site $j$, with periodic boundary conditions imposed as $\sigma_{N+1}^z = \sigma_1^z$. Here, $J_z$ represents the nearest-neighbor coupling strength between spins, and $h_z$ denotes the external magnetic field along the $z$ axis acting on each spin.

With $\hat{H}_S$ and $\hat{H}_E$ defined, we construct the interaction Hamiltonian as
\begin{equation}
    \hat{H}_I = \ketbra{0}\otimes g_0\sum_{j=1}^N\sigma_j^x 
    + \ketbra{1}\otimes g_1\sum_{j=1}^N \sigma_j^y,
\end{equation}
which satisfies the pure dephasing condition in Eq.~\eqref{pure dephasing condition}, while allowing for nontrivial heat dissipation according to Eq.~\eqref{heat dissipation condition}.

We consider a field-dominated weak-coupling regime characterized by $|h_z| > |J_z|$ and $|h_z| \gg g_k$, such that the spin chain remains nearly polarized along the $z$ direction. Furthermore, the parameters are chosen so that thermal excitations are suppressed. In the numerical simulations we fix $h_z = -5\,|J_z|$, $T = |J_z|$, and take $g_0 = g_1 \equiv g$.


\subsection{Decoherence and Non-Markovianity}

Having established the structure of the total system, we now turn to the analysis of its coherence dynamics, focusing on how memory effects emerge from the finite environment. The qubit-environment density operator after the unitary evolution under pure dephasing is given, in the qubit subspace, by
\begin{equation}
    \label{qubit-reservatorio evolved}
    \hat{\rho}_{SE}(t) = \frac{1}{2}\begin{pmatrix}
        \hat{\omega}_0(t)\hat{\rho}_E(0)\,\hat{\omega}_0^\dagger(t) & \hat{\omega}_0(t)\hat{\rho}_E(0)\,\hat{\omega}_1^\dagger(t) \\
        \hat{\omega}_1(t)\hat{\rho}_E(0)\,\hat{\omega}_0^\dagger(t) & \hat{\omega}_1(t)\hat{\rho}_E(0)\,\hat{\omega}_1^\dagger(t) 
    \end{pmatrix}.
\end{equation}
Thus, by tracing out the environment's degrees of freedom, one obtains the reduced state of $S$ as
\begin{equation}\label{reduced system's matrix}
    \hat{\rho}_S(t) = \frac{1}{2}\begin{pmatrix}
        1 &  \Gamma(t) \\
         \Gamma^*(t) & 1 
    \end{pmatrix},
\end{equation}
where $\Gamma(t)$ is defined in Eq.~\eqref{environment coherence} (here the indices are restricted to $k,j = 0,1$ and omitted for simplicity). At this stage, it is straightforward to apply the $l_1$-norm of coherence, which reduces to $C_{l_1}(\hat{\rho}_S(t)) = |\Gamma(t)|$ and encodes all the information about the environment's influence on the system. This resource measure allows us to investigate finite-size effects quantitatively; thus, we compute the time evolution of $|\Gamma(t)|$ numerically for different environmental parameters. 

\begin{figure}[!ht]
    \centering
    \begin{tikzpicture}

  \node[inner sep=0, outer sep=0] (img)
    {\includegraphics[width=0.9\columnwidth]{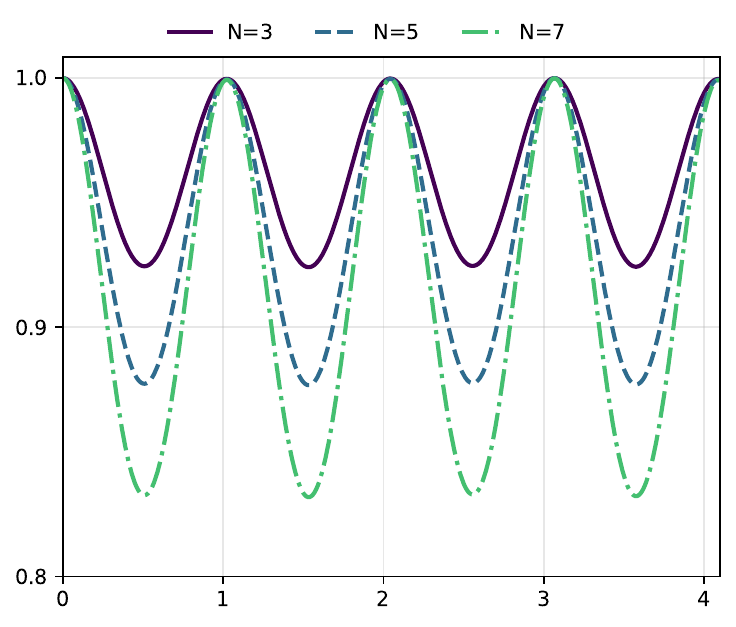}};

  \path[use as bounding box] (img.south west) rectangle (img.north east);

  \begin{scope}[
    shift={(img.south west)},
    x={($(img.south east)-(img.south west)$)},
    y={($(img.north west)-(img.south west)$)}
  ]

    \node[below] at (0.55, 0.03) {$J_z\,t_f$};
    \node[rotate=90] at (-0.02, 0.50) {$|\Gamma(t_f)|$};

  \end{scope}
\end{tikzpicture}
    \caption{
(Color online) Time evolution of the $l_1$-norm of coherence for a qubit coupled to an Ising-like chain in a ring configuration with $g = 0.5\,|J_z|$ for different environment sizes $N = 3, 5, 7$. 
The oscillatory behavior indicates reversible information exchange between the qubit and the finite spin environment.
}
    \label{fig:gamma_var_N}
\end{figure}
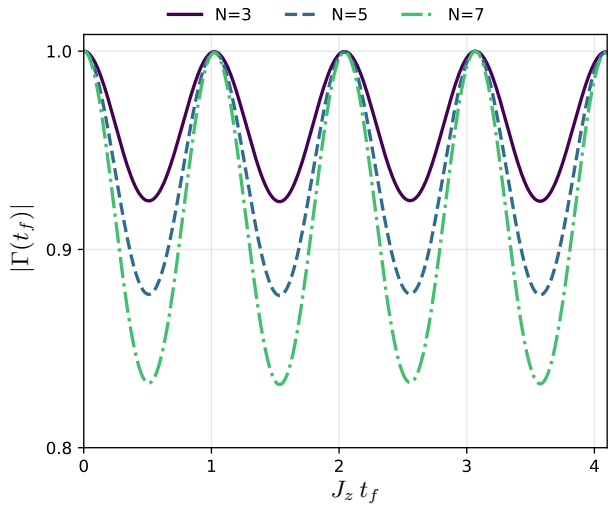

Figure~\ref{fig:gamma_var_N} shows the temporal behavior for different environment sizes $N$ at fixed coupling strength $g$. The dynamics exhibits periodic oscillations of small amplitude, indicating that coherence is nearly preserved in the field-dominated limit. Furthermore, increasing the size of the environment enhances the strength of decoherence. This behavior arises from the growth of the environment's Hilbert space, which increases the number of accessible states and, consequently, the number of transition pathways contributing to the conditional evolution amplitudes.

Additionally, for all chain sizes, the value of $|\Gamma(t)|$ returns to unity after each oscillation period. This behavior corresponds to the reversible exchange of information between the central qubit and the finite spin environment, revealing a non-Markovian character~\cite{mendoncca2024system}.

\begin{figure}[!ht]
    \centering
    \begin{tikzpicture}

  \node[inner sep=0, outer sep=0] (img)
    {\includegraphics[width=0.9\columnwidth]{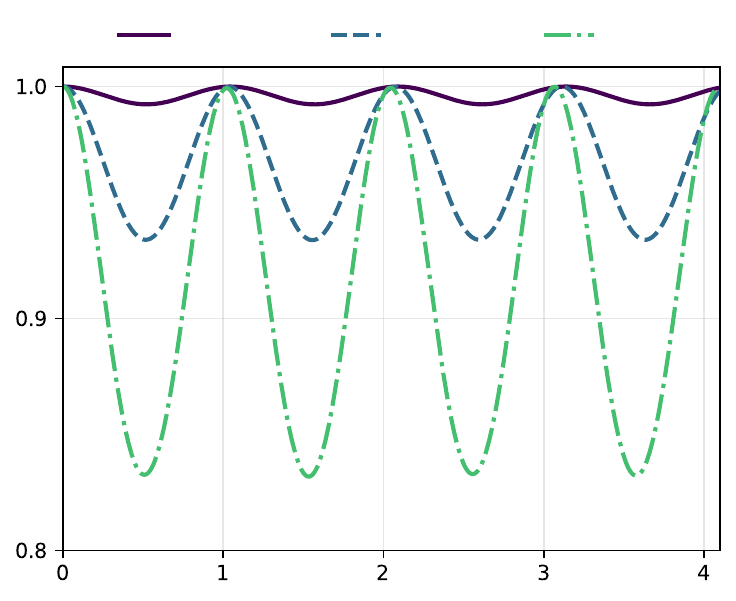}};

  \path[use as bounding box] (img.south west) rectangle (img.north east);

  \begin{scope}[
    shift={(img.south west)},
    x={($(img.south east)-(img.south west)$)},
    y={($(img.north west)-(img.south west)$)}
  ]

    \node[below] at (0.55, 0.03) {$J_z\,t_f$};
    \node[rotate=90] at (-0.02, 0.50) {$|\Gamma(t_f)|$};
    \node[] at (0.31, 0.94){$0.1\,|J_z|$};
    \node[] at (0.6, 0.94){$0.3\,|J_z|$};
    \node[] at (0.89, 0.94){$0.5\,|J_z|$};

  \end{scope}
\end{tikzpicture}
    \caption{
(Color online) Time evolution of the $l_1$-norm of coherence for a qubit coupled to an Ising-like ring environment with fixed size $N = 7$, for different coupling strengths $g = 0.1\,|J_z|$, $0.3\,|J_z|$, and $0.5\,|J_z|$. 
Increasing $g$ enhances the system–environment interaction, leading to stronger decoherence.
}
    \label{fig:gamma_var_g}
\end{figure}
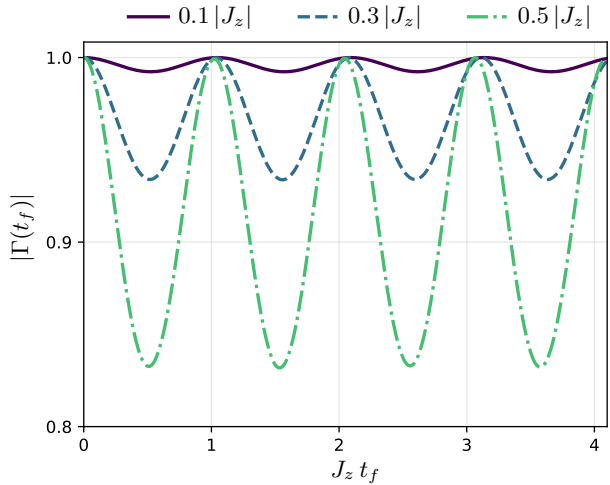

To further examine the influence of the system–environment interaction strength, we analyze the behavior of $|\Gamma(t)|$ for different coupling constants $g$, while keeping the environment size fixed. Figure~\ref{fig:gamma_var_g} shows that increasing the interaction strength enhances the decoherence process, indicating that a larger portion of information becomes temporarily inaccessible to the qubit and that the oscillation period is modified, leading to faster coherence decay. Despite the increased decay for larger values of $g$, well-defined revivals persist for all values of $g$, confirming the non-Markovian character of the dynamics.

Finally, to connect the information backflow between the system and the finite environment, as evidenced by the oscillatory behavior of $|\Gamma(t)|$, we quantify memory effects using the Breuer-Laine-Piilo measure of non-Markovianity, which is based on the trace distance~\cite{breuer2009measure}. The optimal pair that maximizes the growth of distinguishability can be chosen as orthogonal pure states. Thus, we consider two initial states of the central qubit, $\hat{\rho}_S^{\pm}(0) = \ketbra{\psi^{\pm}}$, with $\ket{\psi^{\pm}} = (\ket{0} \pm \ket{1})/\sqrt{2}$, yielding the evolved reduced states as
\begin{equation}
    \label{reduced system's BLP}
    \hat{\rho}_S^{\pm}(t) = \frac{1}{2}\begin{pmatrix}
        1 &  \pm\Gamma(t) \\
         \pm\Gamma^*(t) & 1 
    \end{pmatrix}.
\end{equation}
The trace distance then reads
\begin{equation}
    D(\hat{\rho}_S^+(t), \hat{\rho}_S^-(t)) = \frac{1}{2}||\hat{\rho}_S^+(t) - \hat{\rho}_S^-(t)||_1 = |\Gamma(t)|,
\end{equation}
where $||\hat{A}||_1 = \tr \sqrt{\hat{A}^\dagger\hat{A}}$ is the trace norm, which cannot increase over time for completely positive and trace-preserving maps, reflecting the continuous loss of information, a hallmark of Markovian behavior~\cite{devega2017nonmarkovian}. However, when environmental memory effects are present, the contractivity of the trace distance may break down, and the trace distance can temporarily increase~\cite{breuer2009measure},
\begin{align}
    \sigma_S(t) = \dv{}{t}D(\hat{\rho}_S^+(t), \hat{\rho}_S^-(t)) > 0,
\end{align}
signaling a backflow of information from the environment to the system, a hallmark of non-Markovianity. 

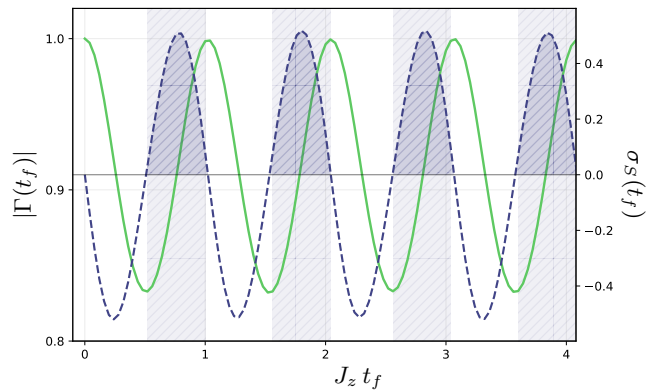
\begin{figure}[!ht]
    \centering
    \begin{tikzpicture}

  \node[inner sep=0, outer sep=0] (img)
    {\includegraphics[width=0.9\columnwidth]{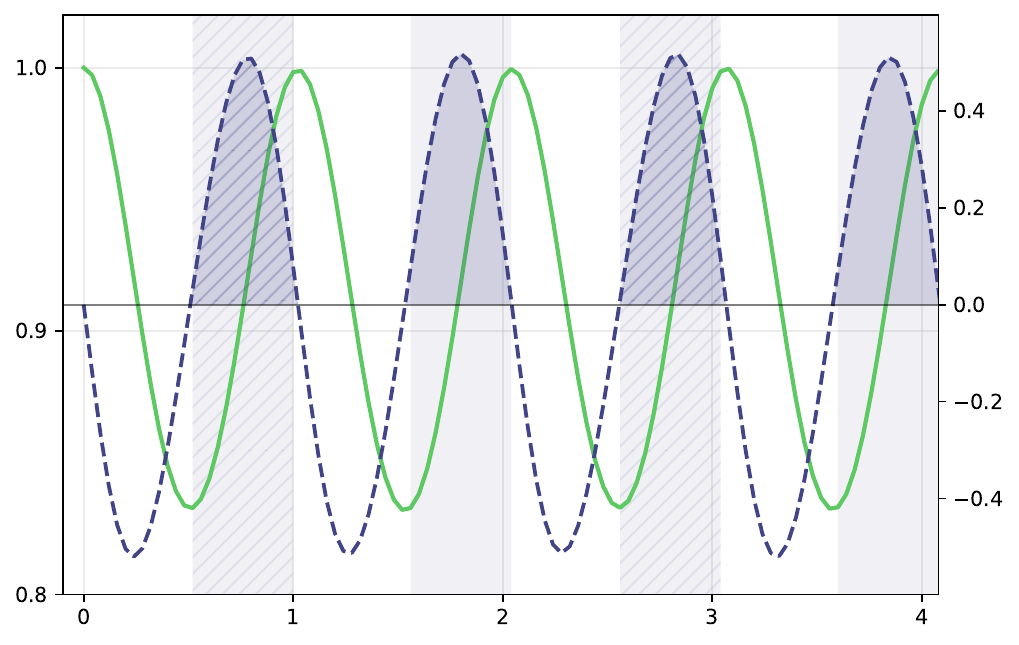}};

  \path[use as bounding box] (img.south west) rectangle (img.north east);

  \begin{scope}[
    shift={(img.south west)},
    x={($(img.south east)-(img.south west)$)},
    y={($(img.north west)-(img.south west)$)}
  ]

    \node[below] at (0.55, 0.03) {$J_z\,t_f$};
    \node[rotate=90] at (-0.02, 0.50) {$|\Gamma(t_f)|$};
    \node[rotate=270] at (1.02, 0.50){$\sigma_S(t_f)$};

  \end{scope}
\end{tikzpicture}
    \caption{
(Color online) Time evolution of the $l_1$-norm of coherence (solid line) and the information flow $\sigma_S(t)$ (dashed line) for $N=7$ and $g=0.5\,|J_z|$. 
Intervals where $\sigma_S(t)$ is positive coincide with coherence revivals, signaling non-Markovian behavior. 
The periodic alternation between decay and revival reflects the finite memory capacity of the spin environment.
}
    \label{fig:non_Markovian}
\end{figure}

Figure~\ref{fig:non_Markovian} illustrates the connection between $|\Gamma(t)|$ and $\sigma_S(t)$ for a fixed chain size $N$ and a fixed coupling strength $g$, showing that positive values of $\sigma_S(t)$ coincide with the time intervals in which $|\Gamma(t)|$ increases. This result confirms that each coherence revival corresponds to a temporary backflow of information from the environment to the system, allowing the system coherence to return close to unity. This behavior reflects the reversible, unitary character of the total dynamics.

\subsection{Heat--Coherence Trade-off and Thermodynamic Interpretation}

The analytical results obtained previously establish that, in the pure dephasing regime, the mean heat dissipated into the environment coincides exactly with the coherent energy contribution in the reformulated first law proposed by Bernardo~\cite{bernardo2020unraveling}, $\langle Q \rangle = \mathcal{C}(t)$. This identity suggests that heat dissipation is not merely an energetic consequence of the interaction, but rather a direct manifestation of coherence dynamics.

To clarify this connection, we note that both the mean heat in Eq.~\eqref{explicit mean heat} and the time-dependent function $\Gamma(t)$ in Eq.~\eqref{environment coherence} depend explicitly on the operators $\hat{\omega}_k(t)$ governing the evolution of the environment conditioned on the central qubit state $\ket{k}$. Therefore, both quantities are governed by the same microscopic mechanism: the noncommutativity between $\hat{H}_E$ and $\hat{\omega}_k(t)$, which induces decoherence, also drives the exchange of energy between the environment and the interaction term. This structural connection implies that the temporal behavior of $\langle Q \rangle$ mirrors the pattern of coherence decay and revival.

\begin{figure}[!ht]
    \centering
    \begin{tikzpicture}

  \node[inner sep=0, outer sep=0] (img)
    {\includegraphics[width=0.9\columnwidth]{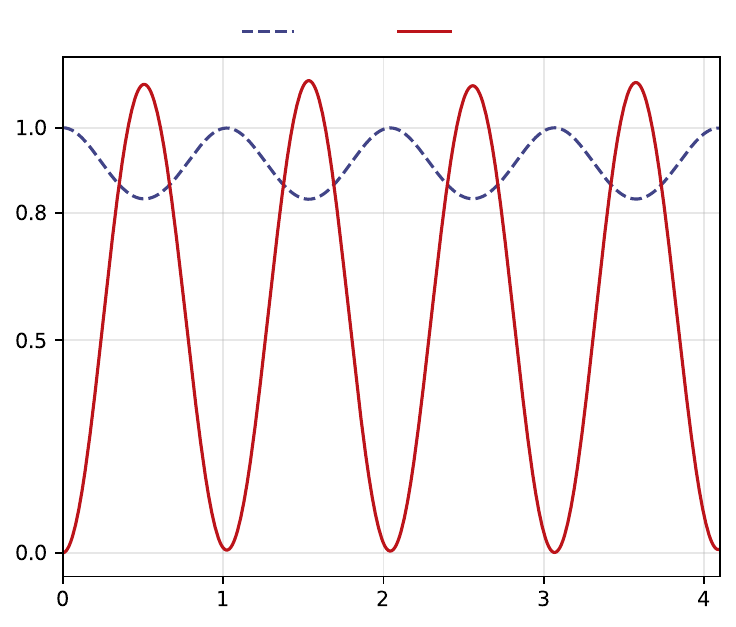}};

  \path[use as bounding box] (img.south west) rectangle (img.north east);

  \begin{scope}[
    shift={(img.south west)},
    x={($(img.south east)-(img.south west)$)},
    y={($(img.north west)-(img.south west)$)}
  ]

    \node[below] at (0.55, 0.03) {$J_z\,t_f$};
    \node[] at (0.46, 0.95){$|\Gamma(t)|$};
    \node[] at (0.66, 0.95){$\langle Q\rangle$};

  \end{scope}
\end{tikzpicture}
    \caption{(Color online) Comparison between the time evolution of the mean heat $\langle Q \rangle$ (solid line) and the $l_1$-norm of coherence (dashed line) for $N=7$ and $g=0.5\,|J_z|$. Each maximum of $\langle Q \rangle$ coincides with a minimum of $|\Gamma(t)|$, indicating that heat dissipation occurs during decoherence.}
    \label{fig:heat_vs_coherence}
\end{figure}
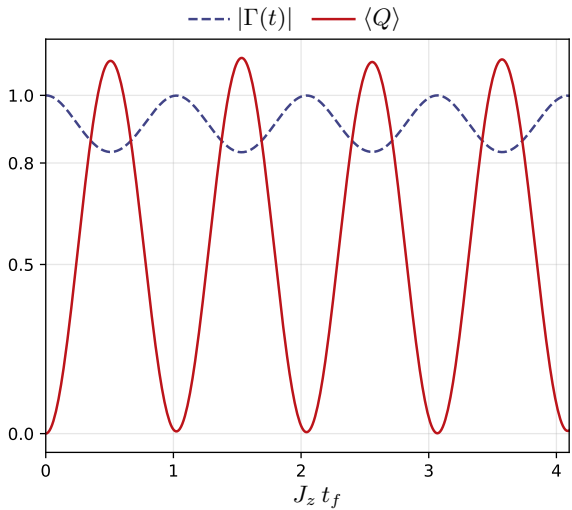

We now investigate numerically how the dissipated heat $\langle Q \rangle$ is linked to the $l_1$-norm of coherence given by $|\Gamma(t)|$, which fully characterizes the decay and revival of the off-diagonal elements of the reduced state $\hat{\rho}_S(t)$. Figure~\ref{fig:heat_vs_coherence} compares these temporal behaviors. 

As can be observed, each peak of $\langle Q \rangle$ corresponds to a minimum of $|\Gamma(t)|$. Conversely, during coherence revivals, the mean heat decreases, indicating that energy is redistributed within the composite system. This intertwined behavior provides numerical evidence that the dissipated heat constitutes a thermodynamic signature of decoherence in composite systems undergoing pure dephasing. When the open system loses coherence, the environment effectively acquires information about it, and when coherence is restored, the information flow reverses.

Altogether, these findings indicate that even in the absence of net energy exchange, the redistribution of coherence into system–environment correlations gives rise to dissipative features.


\section{\label{sec5}Conclusions}

We have investigated the thermodynamic consequences of quantum coherence dynamics in the pure dephasing regime. By combining analytical and numerical methods, we showed that even when the open quantum system exchanges no energy with its environment, nontrivial heat dissipation can arise from the redistribution of energy associated with the interaction term, while the global evolution remains unitary. Within the two-point measurement framework, we established analytically that the mean heat, $\langle Q \rangle$, dissipated into the environment coincides exactly with the coherent energy contribution introduced in Bernardo’s reformulated first law~\cite{bernardo2020unraveling}. This result identifies heat dissipation as a thermodynamic signature of decoherence and demonstrates that decoherence entails an intrinsic energetic cost, even in the absence of system energy exchange, in agreement with Ref.~\cite{popovic2023thermodynamics}.

To exemplify these general results, we analyzed a central qubit coupled to a finite Ising-like spin environment arranged in a ring configuration and quantified its coherence dynamics, heat, and memory effects. The finiteness of the environment induces non-Markovian behavior characterized by information backflow and coherence revivals, which persist across different chain sizes and coupling strengths, reflecting the reversible character of the total unitary dynamics. In particular, our simulations revealed that the mean heat is maximal when coherence is minimal, while coherence revivals are accompanied by reductions in $\langle Q \rangle$. This intertwined behavior indicates that dissipated heat constitutes a thermodynamic signature of decoherence in pure dephasing: when the system loses coherence, the environment effectively acquires information about it, and when coherence is restored, the information flow reverses.

A promising direction for future research is the investigation of entropy production in spin-echo protocols, particularly how it is influenced by the mean heat in the pure dephasing regime. Another important extension involves studying time-dependent coupling strengths to understand how dynamically controlled interactions modify the energetic cost of coupling and decoupling the open quantum system from the environment during pure dephasing dynamics.


\begin{acknowledgments}
We would like to acknowledge T. Debarba and B. L. Bernardo for fruitful discussions and helpful comments. M.P.L. acknowledges financial support from Coordenação de Aperfeiçoamento de Pessoal de Nível Superior - Brasil (CAPES) - Finance Code 001. D.O.S.P. acknowledges support from the Brazilian funding agency CNPq (Grants No. 304891/2022-3 and 402074/2023-8).
\end{acknowledgments}


\bibliography{apssamp}

\providecommand{\noopsort}[1]{}\providecommand{\singleletter}[1]{#1}%
\begin{thebibliography}{30}%
\makeatletter
\providecommand \@ifxundefined [1]{%
 \@ifx{#1\undefined}
}%
\providecommand \@ifnum [1]{%
 \ifnum #1\expandafter \@firstoftwo
 \else \expandafter \@secondoftwo
 \fi
}%
\providecommand \@ifx [1]{%
 \ifx #1\expandafter \@firstoftwo
 \else \expandafter \@secondoftwo
 \fi
}%
\providecommand \natexlab [1]{#1}%
\providecommand \enquote  [1]{``#1''}%
\providecommand \bibnamefont  [1]{#1}%
\providecommand \bibfnamefont [1]{#1}%
\providecommand \citenamefont [1]{#1}%
\providecommand \href@noop [0]{\@secondoftwo}%
\providecommand \href [0]{\begingroup \@sanitize@url \@href}%
\providecommand \@href[1]{\@@startlink{#1}\@@href}%
\providecommand \@@href[1]{\endgroup#1\@@endlink}%
\providecommand \@sanitize@url [0]{\catcode `\\12\catcode `\$12\catcode `\&12\catcode `\#12\catcode `\^12\catcode `\_12\catcode `\%12\relax}%
\providecommand \@@startlink[1]{}%
\providecommand \@@endlink[0]{}%
\providecommand \url  [0]{\begingroup\@sanitize@url \@url }%
\providecommand \@url [1]{\endgroup\@href {#1}{\urlprefix }}%
\providecommand \urlprefix  [0]{URL }%
\providecommand \Eprint [0]{\href }%
\providecommand \doibase [0]{https://doi.org/}%
\providecommand \selectlanguage [0]{\@gobble}%
\providecommand \bibinfo  [0]{\@secondoftwo}%
\providecommand \bibfield  [0]{\@secondoftwo}%
\providecommand \translation [1]{[#1]}%
\providecommand \BibitemOpen [0]{}%
\providecommand \bibitemStop [0]{}%
\providecommand \bibitemNoStop [0]{.\EOS\space}%
\providecommand \EOS [0]{\spacefactor3000\relax}%
\providecommand \BibitemShut  [1]{\csname bibitem#1\endcsname}%
\let\auto@bib@innerbib\@empty
\bibitem [{\citenamefont {Campbell}\ \emph {et~al.}(2026)\citenamefont {Campbell}, \citenamefont {d'Amico}, \citenamefont {Ciampini}, \citenamefont {Anders}, \citenamefont {Ares}, \citenamefont {Artini}, \citenamefont {Auffèves}, \citenamefont {Oftelie}, \citenamefont {Bettmann}, \citenamefont {Bonança} \emph {et~al.}}]{campbell2025roadmap}%
  \BibitemOpen
  \bibfield  {author} {\bibinfo {author} {\bibfnamefont {S.}~\bibnamefont {Campbell}}, \bibinfo {author} {\bibfnamefont {I.}~\bibnamefont {d'Amico}}, \bibinfo {author} {\bibfnamefont {M.~A.}\ \bibnamefont {Ciampini}}, \bibinfo {author} {\bibfnamefont {J.}~\bibnamefont {Anders}}, \bibinfo {author} {\bibfnamefont {N.}~\bibnamefont {Ares}}, \bibinfo {author} {\bibfnamefont {S.}~\bibnamefont {Artini}}, \bibinfo {author} {\bibfnamefont {A.}~\bibnamefont {Auffèves}}, \bibinfo {author} {\bibfnamefont {L.~B.}\ \bibnamefont {Oftelie}}, \bibinfo {author} {\bibfnamefont {L.~P.}\ \bibnamefont {Bettmann}}, \bibinfo {author} {\bibfnamefont {M.~V.}\ \bibnamefont {Bonança}}, \emph {et~al.},\ }\bibfield  {title} {\bibinfo {title} {Roadmap on quantum thermodynamics},\ }\href {https://doi.org/10.1088/2058-9565/ae1e27} {\bibfield  {journal} {\bibinfo  {journal} {Quantum Science and Technology}\ }\textbf {\bibinfo {volume} {11}},\ \bibinfo {pages} {012501} (\bibinfo {year} {2026})}\BibitemShut {NoStop}%
\bibitem [{\citenamefont {Goold}\ \emph {et~al.}(2016)\citenamefont {Goold}, \citenamefont {Huber}, \citenamefont {Riera}, \citenamefont {del Rio},\ and\ \citenamefont {Skrzypczyk}}]{goold2016role}%
  \BibitemOpen
  \bibfield  {author} {\bibinfo {author} {\bibfnamefont {J.}~\bibnamefont {Goold}}, \bibinfo {author} {\bibfnamefont {M.}~\bibnamefont {Huber}}, \bibinfo {author} {\bibfnamefont {A.}~\bibnamefont {Riera}}, \bibinfo {author} {\bibfnamefont {L.}~\bibnamefont {del Rio}},\ and\ \bibinfo {author} {\bibfnamefont {P.}~\bibnamefont {Skrzypczyk}},\ }\bibfield  {title} {\bibinfo {title} {The role of quantum information in thermodynamics: a topical review},\ }\href {https://doi.org/10.1088/1751-8113/49/14/143001} {\bibfield  {journal} {\bibinfo  {journal} {Journal of physics A: mathematical and theoretical}\ }\textbf {\bibinfo {volume} {49}},\ \bibinfo {pages} {143001} (\bibinfo {year} {2016})}\BibitemShut {NoStop}%
\bibitem [{\citenamefont {Hammam}\ \emph {et~al.}(2022)\citenamefont {Hammam}, \citenamefont {Leitch}, \citenamefont {Hassouni},\ and\ \citenamefont {De~Chiara}}]{hammam2022exploiting}%
  \BibitemOpen
  \bibfield  {author} {\bibinfo {author} {\bibfnamefont {K.}~\bibnamefont {Hammam}}, \bibinfo {author} {\bibfnamefont {H.}~\bibnamefont {Leitch}}, \bibinfo {author} {\bibfnamefont {Y.}~\bibnamefont {Hassouni}},\ and\ \bibinfo {author} {\bibfnamefont {G.}~\bibnamefont {De~Chiara}},\ }\bibfield  {title} {\bibinfo {title} {Exploiting coherence for quantum thermodynamic advantage},\ }\href {https://doi.org/10.1088/1367-2630/aca49b} {\bibfield  {journal} {\bibinfo  {journal} {New Journal of Physics}\ }\textbf {\bibinfo {volume} {24}},\ \bibinfo {pages} {113053} (\bibinfo {year} {2022})}\BibitemShut {NoStop}%
\bibitem [{\citenamefont {Su}\ \emph {et~al.}(2018)\citenamefont {Su}, \citenamefont {Chen}, \citenamefont {Ma}, \citenamefont {Chen},\ and\ \citenamefont {Sun}}]{su2018heat}%
  \BibitemOpen
  \bibfield  {author} {\bibinfo {author} {\bibfnamefont {S.}~\bibnamefont {Su}}, \bibinfo {author} {\bibfnamefont {J.}~\bibnamefont {Chen}}, \bibinfo {author} {\bibfnamefont {Y.}~\bibnamefont {Ma}}, \bibinfo {author} {\bibfnamefont {J.}~\bibnamefont {Chen}},\ and\ \bibinfo {author} {\bibfnamefont {C.}~\bibnamefont {Sun}},\ }\bibfield  {title} {\bibinfo {title} {The heat and work of quantum thermodynamic processes with quantum coherence},\ }\href {https://doi.org/10.1088/1674-1056/27/6/060502} {\bibfield  {journal} {\bibinfo  {journal} {Chinese Physics B}\ }\textbf {\bibinfo {volume} {27}},\ \bibinfo {pages} {060502} (\bibinfo {year} {2018})}\BibitemShut {NoStop}%
\bibitem [{\citenamefont {Francica}\ \emph {et~al.}(2019)\citenamefont {Francica}, \citenamefont {Goold},\ and\ \citenamefont {Plastina}}]{francica2019role}%
  \BibitemOpen
  \bibfield  {author} {\bibinfo {author} {\bibfnamefont {G.}~\bibnamefont {Francica}}, \bibinfo {author} {\bibfnamefont {J.}~\bibnamefont {Goold}},\ and\ \bibinfo {author} {\bibfnamefont {F.}~\bibnamefont {Plastina}},\ }\bibfield  {title} {\bibinfo {title} {Role of coherence in the nonequilibrium thermodynamics of quantum systems},\ }\href {https://doi.org/https://doi.org/10.1103/PhysRevE.99.042105} {\bibfield  {journal} {\bibinfo  {journal} {Physical Review E}\ }\textbf {\bibinfo {volume} {99}},\ \bibinfo {pages} {042105} (\bibinfo {year} {2019})}\BibitemShut {NoStop}%
\bibitem [{\citenamefont {Lostaglio}\ \emph {et~al.}(2015)\citenamefont {Lostaglio}, \citenamefont {Jennings},\ and\ \citenamefont {Rudolph}}]{lostaglio2015description}%
  \BibitemOpen
  \bibfield  {author} {\bibinfo {author} {\bibfnamefont {M.}~\bibnamefont {Lostaglio}}, \bibinfo {author} {\bibfnamefont {D.}~\bibnamefont {Jennings}},\ and\ \bibinfo {author} {\bibfnamefont {T.}~\bibnamefont {Rudolph}},\ }\bibfield  {title} {\bibinfo {title} {Description of quantum coherence in thermodynamic processes requires constraints beyond free energy},\ }\href {https://doi.org/https://doi.org/10.1038/ncomms7383} {\bibfield  {journal} {\bibinfo  {journal} {Nature communications}\ }\textbf {\bibinfo {volume} {6}},\ \bibinfo {pages} {6383} (\bibinfo {year} {2015})}\BibitemShut {NoStop}%
\bibitem [{\citenamefont {Korzekwa}\ \emph {et~al.}(2016)\citenamefont {Korzekwa}, \citenamefont {Lostaglio}, \citenamefont {Oppenheim},\ and\ \citenamefont {Jennings}}]{korzekwa2016extraction}%
  \BibitemOpen
  \bibfield  {author} {\bibinfo {author} {\bibfnamefont {K.}~\bibnamefont {Korzekwa}}, \bibinfo {author} {\bibfnamefont {M.}~\bibnamefont {Lostaglio}}, \bibinfo {author} {\bibfnamefont {J.}~\bibnamefont {Oppenheim}},\ and\ \bibinfo {author} {\bibfnamefont {D.}~\bibnamefont {Jennings}},\ }\bibfield  {title} {\bibinfo {title} {The extraction of work from quantum coherence},\ }\href {https://doi.org/10.1088/1367-2630/18/2/023045} {\bibfield  {journal} {\bibinfo  {journal} {New Journal of Physics}\ }\textbf {\bibinfo {volume} {18}},\ \bibinfo {pages} {023045} (\bibinfo {year} {2016})}\BibitemShut {NoStop}%
\bibitem [{\citenamefont {Schlosshauer}(2005)}]{schlosshauer2005decoherence}%
  \BibitemOpen
  \bibfield  {author} {\bibinfo {author} {\bibfnamefont {M.}~\bibnamefont {Schlosshauer}},\ }\bibfield  {title} {\bibinfo {title} {Decoherence, the measurement problem, and interpretations of quantum mechanics},\ }\href {https://doi.org/https://doi.org/10.1103/RevModPhys.76.1267} {\bibfield  {journal} {\bibinfo  {journal} {Reviews of Modern Physics}\ }\textbf {\bibinfo {volume} {76}},\ \bibinfo {pages} {1267} (\bibinfo {year} {2005})}\BibitemShut {NoStop}%
\bibitem [{\citenamefont {Schlosshauer}(2007)}]{schlosshauer2007decoherence}%
  \BibitemOpen
  \bibfield  {author} {\bibinfo {author} {\bibfnamefont {M.}~\bibnamefont {Schlosshauer}},\ }\href {https://doi.org/https://doi.org/10.1007/978-3-540-35775-9_2} {\emph {\bibinfo {title} {Decoherence and the Quantum-to-Classical Transition}}}\ (\bibinfo  {publisher} {Springer},\ \bibinfo {address} {Berlin},\ \bibinfo {year} {2007})\BibitemShut {NoStop}%
\bibitem [{\citenamefont {Zurek}(1991)}]{zurek1991decoherence}%
  \BibitemOpen
  \bibfield  {author} {\bibinfo {author} {\bibfnamefont {W.~H.}\ \bibnamefont {Zurek}},\ }\bibfield  {title} {\bibinfo {title} {Decoherence and the transition from quantum to classical},\ }\href {https://doi.org/https://doi.org/10.1063/1.881293} {\bibfield  {journal} {\bibinfo  {journal} {Physics Today}\ }\textbf {\bibinfo {volume} {44}},\ \bibinfo {pages} {36} (\bibinfo {year} {1991})}\BibitemShut {NoStop}%
\bibitem [{\citenamefont {Rivas}\ and\ \citenamefont {Huelga}(2012)}]{rivas2012open}%
  \BibitemOpen
  \bibfield  {author} {\bibinfo {author} {\bibfnamefont {{\'A}.}~\bibnamefont {Rivas}}\ and\ \bibinfo {author} {\bibfnamefont {S.~F.}\ \bibnamefont {Huelga}},\ }\href@noop {} {\emph {\bibinfo {title} {Open Quantum Systems: An Introduction}}},\ SpringerBriefs in Physics\ (\bibinfo  {publisher} {Springer},\ \bibinfo {address} {Heidelberg},\ \bibinfo {year} {2012})\BibitemShut {NoStop}%
\bibitem [{\citenamefont {Breuer}\ and\ \citenamefont {Petruccione}(2002)}]{breuer2002theory}%
  \BibitemOpen
  \bibfield  {author} {\bibinfo {author} {\bibfnamefont {H.-P.}\ \bibnamefont {Breuer}}\ and\ \bibinfo {author} {\bibfnamefont {F.}~\bibnamefont {Petruccione}},\ }\href@noop {} {\emph {\bibinfo {title} {The theory of open quantum systems}}}\ (\bibinfo  {publisher} {OUP Oxford},\ \bibinfo {year} {2002})\BibitemShut {NoStop}%
\bibitem [{\citenamefont {Joos}\ and\ \citenamefont {Zeh}(1985)}]{joos1985emergence}%
  \BibitemOpen
  \bibfield  {author} {\bibinfo {author} {\bibfnamefont {E.}~\bibnamefont {Joos}}\ and\ \bibinfo {author} {\bibfnamefont {H.~D.}\ \bibnamefont {Zeh}},\ }\bibfield  {title} {\bibinfo {title} {The emergence of classical properties through interaction with the environment},\ }\href {https://doi.org/https://doi.org/10.1007/BF01725541} {\bibfield  {journal} {\bibinfo  {journal} {Zeitschrift f{\"u}r Physik B Condensed Matter}\ }\textbf {\bibinfo {volume} {59}},\ \bibinfo {pages} {223} (\bibinfo {year} {1985})}\BibitemShut {NoStop}%
\bibitem [{\citenamefont {Chen}\ \emph {et~al.}(2019)\citenamefont {Chen}, \citenamefont {Lo}, \citenamefont {Gneiting}, \citenamefont {Bae}, \citenamefont {Chen},\ and\ \citenamefont {Nori}}]{chen2019quantifying}%
  \BibitemOpen
  \bibfield  {author} {\bibinfo {author} {\bibfnamefont {H.-B.}\ \bibnamefont {Chen}}, \bibinfo {author} {\bibfnamefont {P.-Y.}\ \bibnamefont {Lo}}, \bibinfo {author} {\bibfnamefont {C.}~\bibnamefont {Gneiting}}, \bibinfo {author} {\bibfnamefont {J.}~\bibnamefont {Bae}}, \bibinfo {author} {\bibfnamefont {Y.-N.}\ \bibnamefont {Chen}},\ and\ \bibinfo {author} {\bibfnamefont {F.}~\bibnamefont {Nori}},\ }\bibfield  {title} {\bibinfo {title} {Quantifying the nonclassicality of pure dephasing},\ }\href {https://doi.org/https://doi.org/10.1038/s41467-019-11502-4} {\bibfield  {journal} {\bibinfo  {journal} {Nature communications}\ }\textbf {\bibinfo {volume} {10}},\ \bibinfo {pages} {3794} (\bibinfo {year} {2019})}\BibitemShut {NoStop}%
\bibitem [{\citenamefont {Roszak}\ and\ \citenamefont {Cywi{\'n}ski}(2015{\natexlab{a}})}]{roszak2015qubit}%
  \BibitemOpen
  \bibfield  {author} {\bibinfo {author} {\bibfnamefont {K.}~\bibnamefont {Roszak}}\ and\ \bibinfo {author} {\bibfnamefont {{\L}.}~\bibnamefont {Cywi{\'n}ski}},\ }\bibfield  {title} {\bibinfo {title} {Qubit-environment entanglement generation and the spin echo},\ }\href {https://doi.org/https://doi.org/10.1103/PhysRevA.103.032208} {\bibfield  {journal} {\bibinfo  {journal} {Physical Review A}\ }\textbf {\bibinfo {volume} {92}},\ \bibinfo {pages} {032310} (\bibinfo {year} {2015}{\natexlab{a}})}\BibitemShut {NoStop}%
\bibitem [{\citenamefont {Roszak}\ and\ \citenamefont {Cywi{\'n}ski}(2015{\natexlab{b}})}]{roszak2015characterization}%
  \BibitemOpen
  \bibfield  {author} {\bibinfo {author} {\bibfnamefont {K.}~\bibnamefont {Roszak}}\ and\ \bibinfo {author} {\bibfnamefont {{\L}.}~\bibnamefont {Cywi{\'n}ski}},\ }\bibfield  {title} {\bibinfo {title} {Characterization and measurement of qubit-environment-entanglement generation during pure dephasing},\ }\href {https://doi.org/https://doi.org/10.1103/PhysRevA.92.032310} {\bibfield  {journal} {\bibinfo  {journal} {Physical Review A}\ }\textbf {\bibinfo {volume} {92}},\ \bibinfo {pages} {032310} (\bibinfo {year} {2015}{\natexlab{b}})}\BibitemShut {NoStop}%
\bibitem [{\citenamefont {Krovi}\ \emph {et~al.}(2007)\citenamefont {Krovi}, \citenamefont {Oreshkov}, \citenamefont {Ryazanov},\ and\ \citenamefont {Lidar}}]{krovi2007nonmarkovian}%
  \BibitemOpen
  \bibfield  {author} {\bibinfo {author} {\bibfnamefont {H.}~\bibnamefont {Krovi}}, \bibinfo {author} {\bibfnamefont {O.}~\bibnamefont {Oreshkov}}, \bibinfo {author} {\bibfnamefont {M.}~\bibnamefont {Ryazanov}},\ and\ \bibinfo {author} {\bibfnamefont {D.~A.}\ \bibnamefont {Lidar}},\ }\bibfield  {title} {\bibinfo {title} {Non-markovian dynamics of a qubit coupled to an ising spin bath},\ }\href {https://doi.org/https://doi.org/10.1103/PhysRevA.76.052117} {\bibfield  {journal} {\bibinfo  {journal} {Physical Review A}\ }\textbf {\bibinfo {volume} {76}},\ \bibinfo {pages} {052117} (\bibinfo {year} {2007})}\BibitemShut {NoStop}%
\bibitem [{\citenamefont {Xiong}\ \emph {et~al.}(2020)\citenamefont {Xiong}, \citenamefont {Li},\ and\ \citenamefont {Chen}}]{xiong2020boson}%
  \BibitemOpen
  \bibfield  {author} {\bibinfo {author} {\bibfnamefont {F.-L.}\ \bibnamefont {Xiong}}, \bibinfo {author} {\bibfnamefont {L.}~\bibnamefont {Li}},\ and\ \bibinfo {author} {\bibfnamefont {Z.-B.}\ \bibnamefont {Chen}},\ }\bibfield  {title} {\bibinfo {title} {Boson-boson pure-dephasing model with non-markovian properties},\ }\href {https://doi.org/https://doi.org/10.1016/j.physleta.2018.10.022} {\bibfield  {journal} {\bibinfo  {journal} {Physical Review A}\ }\textbf {\bibinfo {volume} {101}},\ \bibinfo {pages} {062102} (\bibinfo {year} {2020})}\BibitemShut {NoStop}%
\bibitem [{\citenamefont {Dubertrand}\ \emph {et~al.}(2018)\citenamefont {Dubertrand}, \citenamefont {Cesa},\ and\ \citenamefont {Martin}}]{dubertrand2018analytical}%
  \BibitemOpen
  \bibfield  {author} {\bibinfo {author} {\bibfnamefont {R.}~\bibnamefont {Dubertrand}}, \bibinfo {author} {\bibfnamefont {A.}~\bibnamefont {Cesa}},\ and\ \bibinfo {author} {\bibfnamefont {J.}~\bibnamefont {Martin}},\ }\bibfield  {title} {\bibinfo {title} {Analytical results for the quantum non-markovianity of spin ensembles undergoing pure dephasing dynamics},\ }\href {https://doi.org/https://doi.org/10.1103/PhysRevA.97.062126} {\bibfield  {journal} {\bibinfo  {journal} {Physical Review A}\ }\textbf {\bibinfo {volume} {97}},\ \bibinfo {pages} {062126} (\bibinfo {year} {2018})}\BibitemShut {NoStop}%
\bibitem [{\citenamefont {Popovic}\ \emph {et~al.}(2023)\citenamefont {Popovic}, \citenamefont {Mitchison},\ and\ \citenamefont {Goold}}]{popovic2023thermodynamics}%
  \BibitemOpen
  \bibfield  {author} {\bibinfo {author} {\bibfnamefont {M.}~\bibnamefont {Popovic}}, \bibinfo {author} {\bibfnamefont {M.~T.}\ \bibnamefont {Mitchison}},\ and\ \bibinfo {author} {\bibfnamefont {J.}~\bibnamefont {Goold}},\ }\bibfield  {title} {\bibinfo {title} {Thermodynamics of decoherence},\ }\href {https://doi.org/10.1098/rspa.2023.0040} {\bibfield  {journal} {\bibinfo  {journal} {Proceedings of the Royal Society A}\ }\textbf {\bibinfo {volume} {479}},\ \bibinfo {pages} {20230040} (\bibinfo {year} {2023})}\BibitemShut {NoStop}%
\bibitem [{\citenamefont {Bernardo}(2020)}]{bernardo2020unraveling}%
  \BibitemOpen
  \bibfield  {author} {\bibinfo {author} {\bibfnamefont {B.~d.~L.}\ \bibnamefont {Bernardo}},\ }\bibfield  {title} {\bibinfo {title} {Unraveling the role of coherence in the first law of quantum thermodynamics},\ }\href {https://doi.org/https://doi.org/10.1103/PhysRevE.102.062152} {\bibfield  {journal} {\bibinfo  {journal} {Physical Review E}\ }\textbf {\bibinfo {volume} {102}},\ \bibinfo {pages} {062152} (\bibinfo {year} {2020})}\BibitemShut {NoStop}%
\bibitem [{\citenamefont {Baumgratz}\ \emph {et~al.}(2014)\citenamefont {Baumgratz}, \citenamefont {Cramer},\ and\ \citenamefont {Plenio}}]{baumgratz2013quantifying}%
  \BibitemOpen
  \bibfield  {author} {\bibinfo {author} {\bibfnamefont {T.}~\bibnamefont {Baumgratz}}, \bibinfo {author} {\bibfnamefont {M.}~\bibnamefont {Cramer}},\ and\ \bibinfo {author} {\bibfnamefont {M.~B.}\ \bibnamefont {Plenio}},\ }\bibfield  {title} {\bibinfo {title} {Quantifying coherence},\ }\href {https://doi.org/10.1103/PhysRevLett.113.140401} {\bibfield  {journal} {\bibinfo  {journal} {Phys. Rev. Lett.}\ }\textbf {\bibinfo {volume} {113}},\ \bibinfo {pages} {140401} (\bibinfo {year} {2014})}\BibitemShut {NoStop}%
\bibitem [{\citenamefont {Filippetto}\ \emph {et~al.}(2025)\citenamefont {Filippetto}, \citenamefont {Costa},\ and\ \citenamefont {Zawadzki}}]{filippetto2025convite}%
  \BibitemOpen
  \bibfield  {author} {\bibinfo {author} {\bibfnamefont {M.~E.~R.}\ \bibnamefont {Filippetto}}, \bibinfo {author} {\bibfnamefont {A.~C.~S.}\ \bibnamefont {Costa}},\ and\ \bibinfo {author} {\bibfnamefont {K.}~\bibnamefont {Zawadzki}},\ }\bibfield  {title} {\bibinfo {title} {Um convite {\`a} termodin{\^a}mica qu{\^a}ntica},\ }\href {https://doi.org/https://doi.org/10.1590/1806-9126-RBEF-2025-0391} {\bibfield  {journal} {\bibinfo  {journal} {Revista Brasileira de Ensino de F{\'\i}sica}\ }\textbf {\bibinfo {volume} {47}},\ \bibinfo {pages} {e20250391} (\bibinfo {year} {2025})}\BibitemShut {NoStop}%
\bibitem [{\citenamefont {Talkner}\ \emph {et~al.}(2007)\citenamefont {Talkner}, \citenamefont {Lutz},\ and\ \citenamefont {H{\"a}nggi}}]{talkner2007fluctuation}%
  \BibitemOpen
  \bibfield  {author} {\bibinfo {author} {\bibfnamefont {P.}~\bibnamefont {Talkner}}, \bibinfo {author} {\bibfnamefont {E.}~\bibnamefont {Lutz}},\ and\ \bibinfo {author} {\bibfnamefont {P.}~\bibnamefont {H{\"a}nggi}},\ }\bibfield  {title} {\bibinfo {title} {Fluctuation theorems: Work is not an observable},\ }\href {https://doi.org/https://doi.org/10.1103/PhysRevE.75.050102} {\bibfield  {journal} {\bibinfo  {journal} {Physical Review E}\ }\textbf {\bibinfo {volume} {75}},\ \bibinfo {pages} {050102} (\bibinfo {year} {2007})}\BibitemShut {NoStop}%
\bibitem [{\citenamefont {Talkner}\ and\ \citenamefont {H{\"a}nggi}(2020)}]{talkner2020colloquium}%
  \BibitemOpen
  \bibfield  {author} {\bibinfo {author} {\bibfnamefont {P.}~\bibnamefont {Talkner}}\ and\ \bibinfo {author} {\bibfnamefont {P.}~\bibnamefont {H{\"a}nggi}},\ }\bibfield  {title} {\bibinfo {title} {Colloquium: Statistical mechanics and thermodynamics at strong coupling: Quantum and classical},\ }\href {https://doi.org/https://doi.org/10.1103/RevModPhys.92.041002} {\bibfield  {journal} {\bibinfo  {journal} {Reviews of Modern Physics}\ }\textbf {\bibinfo {volume} {92}},\ \bibinfo {pages} {041002} (\bibinfo {year} {2020})}\BibitemShut {NoStop}%
\bibitem [{\citenamefont {Esposito}\ \emph {et~al.}(2010)\citenamefont {Esposito}, \citenamefont {Lindenberg},\ and\ \citenamefont {Van~den Broeck}}]{esposito2010entropy}%
  \BibitemOpen
  \bibfield  {author} {\bibinfo {author} {\bibfnamefont {M.}~\bibnamefont {Esposito}}, \bibinfo {author} {\bibfnamefont {K.}~\bibnamefont {Lindenberg}},\ and\ \bibinfo {author} {\bibfnamefont {C.}~\bibnamefont {Van~den Broeck}},\ }\bibfield  {title} {\bibinfo {title} {Entropy production as correlation between system and reservoir},\ }\href {https://doi.org/10.1088/1367-2630/12/1/013013} {\bibfield  {journal} {\bibinfo  {journal} {New Journal of Physics}\ }\textbf {\bibinfo {volume} {12}},\ \bibinfo {pages} {013013} (\bibinfo {year} {2010})}\BibitemShut {NoStop}%
\bibitem [{\citenamefont {Talkner}\ \emph {et~al.}(2016)\citenamefont {Talkner}, \citenamefont {Campisi},\ and\ \citenamefont {H{\"a}nggi}}]{talkner2016fluctuation}%
  \BibitemOpen
  \bibfield  {author} {\bibinfo {author} {\bibfnamefont {P.}~\bibnamefont {Talkner}}, \bibinfo {author} {\bibfnamefont {M.}~\bibnamefont {Campisi}},\ and\ \bibinfo {author} {\bibfnamefont {P.}~\bibnamefont {H{\"a}nggi}},\ }\bibfield  {title} {\bibinfo {title} {Fluctuation theorems in driven open quantum systems},\ }\href {https://doi.org/10.1088/1742-5468/2009/02/P02025} {\bibfield  {journal} {\bibinfo  {journal} {Journal of Statistical Mechanics: Theory and Experiment}\ }\textbf {\bibinfo {volume} {2016}},\ \bibinfo {pages} {114004} (\bibinfo {year} {2016})}\BibitemShut {NoStop}%
\bibitem [{\citenamefont {Mendon{\c{c}}a}\ \emph {et~al.}(2024)\citenamefont {Mendon{\c{c}}a}, \citenamefont {C{\'e}leri}, \citenamefont {Paternostro},\ and\ \citenamefont {Soares-Pinto}}]{mendoncca2024system}%
  \BibitemOpen
  \bibfield  {author} {\bibinfo {author} {\bibfnamefont {T.~M.}\ \bibnamefont {Mendon{\c{c}}a}}, \bibinfo {author} {\bibfnamefont {L.~C.}\ \bibnamefont {C{\'e}leri}}, \bibinfo {author} {\bibfnamefont {M.}~\bibnamefont {Paternostro}},\ and\ \bibinfo {author} {\bibfnamefont {D.~O.}\ \bibnamefont {Soares-Pinto}},\ }\bibfield  {title} {\bibinfo {title} {System-environment quantum information flow},\ }\href {https://doi.org/https://doi.org/10.1103/PhysRevA.110.L040401} {\bibfield  {journal} {\bibinfo  {journal} {Physical Review A}\ }\textbf {\bibinfo {volume} {110}},\ \bibinfo {pages} {L040401} (\bibinfo {year} {2024})}\BibitemShut {NoStop}%
\bibitem [{\citenamefont {Breuer}\ \emph {et~al.}(2009)\citenamefont {Breuer}, \citenamefont {Laine},\ and\ \citenamefont {Piilo}}]{breuer2009measure}%
  \BibitemOpen
  \bibfield  {author} {\bibinfo {author} {\bibfnamefont {H.-P.}\ \bibnamefont {Breuer}}, \bibinfo {author} {\bibfnamefont {E.-M.}\ \bibnamefont {Laine}},\ and\ \bibinfo {author} {\bibfnamefont {J.}~\bibnamefont {Piilo}},\ }\bibfield  {title} {\bibinfo {title} {Measure for the degree of non-markovian behavior of quantum processes in open systems},\ }\href {https://doi.org/10.1103/PhysRevLett.103.210401} {\bibfield  {journal} {\bibinfo  {journal} {Physical Review Letters}\ }\textbf {\bibinfo {volume} {103}},\ \bibinfo {pages} {210401} (\bibinfo {year} {2009})}\BibitemShut {NoStop}%
\bibitem [{\citenamefont {de~Vega}\ and\ \citenamefont {Alonso}(2017)}]{devega2017nonmarkovian}%
  \BibitemOpen
  \bibfield  {author} {\bibinfo {author} {\bibfnamefont {I.}~\bibnamefont {de~Vega}}\ and\ \bibinfo {author} {\bibfnamefont {D.}~\bibnamefont {Alonso}},\ }\bibfield  {title} {\bibinfo {title} {Dynamics of non-markovian open quantum systems},\ }\href {https://doi.org/https://doi.org/10.1103/RevModPhys.89.015001} {\bibfield  {journal} {\bibinfo  {journal} {Reviews of Modern Physics}\ }\textbf {\bibinfo {volume} {89}},\ \bibinfo {pages} {015001} (\bibinfo {year} {2017})}\BibitemShut {NoStop}%
\end{thebibliography}%

\end{document}